\documentclass[12pt,tightenlines,eqsecnum,aps,amsmath,amssymb,nofootinbib,prd]{revtex4}

\usepackage{graphicx}% Include figure files
\usepackage{dcolumn}% Align table columns on decimal point
\usepackage{bm}% bold math
\usepackage{dsfont} %for symbols of Real Complex....

\usepackage{tikz}
\usetikzlibrary{positioning}
\begin{document}

%\preprint{APS/123-QED}

\title{Non-oscillatory power spectrum from States of Low Energy in kinetically dominated early universes}% Force line breaks with \\

\author{Mercedes Mart\'in-Benito}
 \email{m.martin.benito@ucm.es}
\author{Rita B. Neves}%
 \email{rneves@ucm.es}
\affiliation{Departamento de F\'isica Te\'orica and IPARCOS, Universidad Complutense de Madrid, Parque de Ciencias 1, 28040 Madrid, Spain}

\author{Javier Olmedo}
 \email{javolmedo@ugr.es}
\affiliation{Departamento de F\'isica Te\'orica y del Cosmos, Universidad de Granada, 18071 Granada, Spain}%

\date{\today}% It is always \today, today,

\begin{abstract}
Recently, States of Low Energy (SLEs) have been proposed as viable vacuum states of primordial perturbations within Loop Quantum Cosmology (LQC). In this work we investigate the effect of the high curvature region of LQC on the definition of SLEs. Shifting the support of the test function that defines them away from this regime results in primordial power spectra of perturbations closer to those of the  so-called Non-oscillatory (NO) vacuum, which is another viable choice of initial conditions previously introduced in the LQC context. Furthermore, through a comparison with the Hadamard-like SLEs, we prove that the NO vacuum is of Hadamard type as well.
\end{abstract}

\maketitle

%\tableofcontents

\section{Introduction}

In a previous work \cite{SLEs_ours}, we have proposed the States of Low Energy (SLEs) introduced in \cite{Olbermann2007} as viable candidates for the vacuum state of cosmological perturbations in Loop Quantum Cosmology (LQC). We were motivated by the fact that they were proven to be Hadamard states that minimized the regularized energy density when smeared along the time-like curve of an isotropic observer via a test function. Furthermore, they had been shown to provide a qualitative behavior in the ultraviolet (UV) and infrared regimes of the primordial power spectra of scalar and tensor perturbations that agrees with observations in models where a period of kinetic dominance precedes inflation \cite{Niedermaier2020}, which is the case in LQC. However, in \cite{SLEs_ours} we have only considered test functions that could be seen as natural choices within LQC, namely, ones with support on the high curvature regime. As long as this is the case, we have shown that the ambiguity in the introduction of the test function is surpassed in this context, in the sense that the resulting SLE and power spectra seem to be very insensitive to its shape and support, provided it is wide enough.

In this work, we investigate the effect of shifting the test function away from the high curvature regime. Firstly, this provides a more complete analysis of the SLEs and the ambiguity of the test function. Secondly, this allows us to distinguish in the primordial power spectra the consequences coming directly from LQC corrections and those related to having a period of kinetic dominance prior to inflation, which can also be obtained in a classical scenario. We will show that if the test function ignores the Planckian region, the effect in the resulting SLE is appreciable. Furthermore, in the power spectra the oscillations that were previously found for lower wave numbers are now dampened. 

This motivates us to compare our results with those found in the LQC literature that adopts as initial conditions for the perturbations the so-called non-oscillatory (NO) vacuum state \cite{nonosc,hyb-obs}. As the name suggests, this state is precisely defined to minimize mode by mode the amplitude of the oscillations of the primordial power spectra in a given time interval. It turns out that this minimization in time is reflected in a minimization of oscillations in the $k$ domain of the power spectra. This NO prescription has been motivated as well as a good candidate for the vacuum of the perturbations \cite{menava}. One question that so far remained unanswered is whether this NO vacuum is or not of Hadamard type. In this work, by comparing with SLEs, we show that indeed this is the case. To do so, we resort to their UV expansions, obtained for SLEs in \cite{Niedermaier2020} and for the NO vacuum in \cite{menava}.

This manuscript is organized as follows. In Section \ref{sec:SLEs in LQC} we review the application of SLEs in LQC as presented in \cite{SLEs_ours}. In Section \ref{sec:SLEs no bounce} we explore the consequence of excluding the high curvature regime from the test function, computing the corresponding SLEs and power spectra at the end of inflation. Section \ref{sec:SLEvsNO} is devoted to a proof that the NO vacuum is Hadamard, based on the comparison with SLEs in the UV limit. Finally, we conclude in Section \ref{sec:conc} with a discussion and closing remarks.

Throughout we adopt Planck units $c=\hbar=G=1$ for numerical computations, keeping factors of $G$ in expressions.

\section{Cosmological perturbations and States of Low Energy in LQC}\label{sec:SLEs in LQC}

In this section we will briefly review the dynamics of cosmological perturbations in LQC through its hybrid approach, as well as the definition of SLEs in this context, as exposed in \cite{SLEs_ours}. Let us start by considering the spatially flat FLRW model with scale factor $a$, minimally coupled to the scalar field $\phi$ subject to the potential $V(\phi)$, which will drive inflation. Cosmological perturbations are usually described by scalar and tensor gauge invariant perturbations $\mathcal{Q}$ and $\mathcal{T}^{I}$ respectively, where $I$ denotes the two possible polarizations of tensor perturbations. Expanding in Fourier modes $\mathcal{Q}_k$ and $\mathcal{T}^{I}_k$, we can write the equation of motion for each mode with wave number $k=|\vec{k}|$ as
\begin{align}
    \ddot{\mathcal{Q}}_k + 3H(t) \dot{\mathcal{Q}}_k + \left(\omega^{(s)}_k(t)\right)^2 \mathcal{Q}_k = 0,\label{eq:eom_Qk}\\
    \ddot{\mathcal{T}}^{I}_k + 3H(t) \dot{\mathcal{T}}^{I}_k + \left(\omega^{(t)}_k(t)\right)^2 \mathcal{T}^{I}_k = 0,\label{eq:eom_Tk}
\end{align}
where the dot denotes derivative with respect to cosmological time $t$, and $H= \dot{a}/a$ is the Hubble parameter. As we will discuss further ahead, the form of the terms $\omega_k$ depends on the quantization. It is common to work with the rescaled fields $u = a \mathcal{Q}$, $\mu^{I} = a \mathcal{T}^{I}$, and in conformal time $\eta$, such that $d\eta=dt/a$. Then, we find the equations of motion of the Fourier modes of these fields, $u_k$ and $\mu_k$ respectively, to be
\begin{align}
    u^{\prime\prime}_{k}(\eta) + \left(k^2+s^{(s)}(\eta)\right) u_{k}(\eta) &= 0,\label{eq:eom_uk}\\
    \left(\mu^I_k(\eta)\right)^{\prime\prime} + \left(k^2+s^{(t)}(\eta)\right) \mu^I_{k}(\eta) &= 0,\label{eq:eom_muk}
\end{align}
where the prime denotes derivative with respect to conformal time $\eta$ and $s^{(s)}(\eta)$ and $s^{(t)}(\eta)$ are the time-dependent mass terms of scalar and tensor modes respectively. From the hybrid approach to LQC, one can write these as functions of the background variables $a$, $\rho$ (inflaton energy density), $P$ (inflaton pressure) and the inflaton potential $V(\phi)$ as \cite{hyb-vs-dress}:
\begin{equation}
    s^{(t)} = -\frac{4\pi G}{3} a^2 \left(\rho - 3P\right),\qquad s^{(s)} =  s^{(t)} + \mathcal{U},
\end{equation}
where
\begin{equation}\label{eq:U_MS}
    \mathcal{U} = a^2\left[V_{,\phi\phi}+48\pi G V(\phi)+ 6 \frac{a^{\prime}\phi^{\prime}}{a^3 \rho} V_{,\phi}-\frac{48\pi G}{\rho} V^2(\phi)\right].
\end{equation}
To simplify notation, in the following we will use $s(\eta)$ to refer generically to both of them, as our comments apply equally to both scalar and tensor modes. When doing so, for simplicity, we will refer only to $u$ as everything is analogous for $\mu^{I}$. It is easy to find that $s(\eta)$ can be related to $\omega^2_k$, now written in terms of conformal time, through
\begin{equation}
    \omega^2_k(\eta) = \frac{1}{a^2(\eta)}\left[k^2 + s(\eta) + \frac{a^{\prime\prime}(\eta)}{a(\eta)}\right].
\end{equation}

Generally, there are no analytical solutions to such equations of motion, and results have to be obtained numerically, given initial conditions $u_k(0)$, $u'_k(0)$. These can be parametrized up to a phase through
\begin{equation}\label{eq:CkDk}
    u_k(0) = \frac{1}{\sqrt{2 D_k}}, \qquad u_k^{\prime}(0) = \sqrt{\frac{D_k}{2}}\left(C_k-i\right),
\end{equation}
where $D_k$ is a positive function and $C_k$ any real function. Once defined, the perturbations can be evolved until a time $\eta_{\rm end}$ during inflation when all the scales of interest have crossed the horizon. The primordial power spectra of the comoving curvature perturbation $\mathcal{R}_k=u_{k}/z$ (where $z=a\dot{\phi}/H$) and tensor perturbations $\mathcal{T}$, defined as
\begin{equation}
    \mathcal{P}_{\mathcal{R}}(k) = \frac{k^{3}}{2\pi^{2}}\frac{|u_{k}|^{2}}{z^{2}}, \qquad \mathcal{P}_{\mathcal{T}}(k) = \frac{32k^{3}}{\pi}\frac{|\mu_{k}^{I}|^{2}}{a^{2}},\label{eq:PS_RT}
\end{equation}
are evaluated at $\eta = \eta_{\rm end}$. The choice of initial conditions amounts to a choice of vacuum state for the perturbations. In this context, there is no notion of a unique natural vacuum. Indeed, several proposals have been made of initial vacua within the LQC framework \cite{nonosc,menava,instvac,Ashtekar:2016} that result in primordial power spectra compatible with observations. In these analyses, initial conditions are set at the LQC bounce where the scale factor of the geometry reaches a minimum, and then it starts expanding. At this bounce, the spacetime curvature reaches a maximum value of the order of the Planck scale.  The work of \cite{SLEs_ours} applied the SLE construction defined in \cite{Olbermann2007} to this context. These are defined as the states that minimize the energy density smeared along a time-like curve, specified by a test function $f$. In the following we summarize this procedure, adapted to our notation (namely working with $u$ and $\mu$ and in conformal time). For further details we refer the reader to \cite{SLEs_ours,Olbermann2007}. Given a fiducial solution $v$ to the equation of motion \eqref{eq:eom_uk}, the SLE associated to the test function $f(\eta)$ is found through the Bogoliubov transformation
\begin{equation}\label{eq:bogo v to u}
    u_k = \alpha(k) v_k + \beta(k) \bar{v}_k,
\end{equation}
where the Bogoliubov coefficients $\alpha(k)$ and $\beta(k)$ are found uniquely (up to a phase) to be
\begin{align}
    \beta(k) &= \sqrt{\frac{c_1(k)}{2\sqrt{c_1^2(k)-|c_2^2(k)|}}-\frac{1}{2}}\ ,\label{eq:beta}\\
    \alpha(k) &= -e^{-i\text{Arg}[c_2(k)]} \sqrt{\frac{c_1(k)}{2\sqrt{c_1^2(k)-|c_2^2(k)|}}+\frac{1}{2}}\ .\label{eq:alpha}
\end{align}
with
\begin{align}
    c_1(k) &:= \frac{1}{2} \int d\eta\,f^2(\eta) a \left[\left\lvert\left(\frac{v_k}{a}\right)^{\prime}\right\rvert^2 +\omega_k^2 \left\lvert\frac{v_k}{a}\right\rvert^2\right],\label{eq:c1}\\
    c_2(k) &:= \frac{1}{2} \int d\eta\,f^2(\eta) a \left[\left(\left(\frac{v_k}{a}\right)^{\prime}\right)^2 +\omega_k^2\ \frac{v^2_k}{a^2}\right],\label{eq:c2}
\end{align}
Note that these quantities carry a dependence on the test function $f$. Indeed, as remarked, equation \eqref{eq:bogo v to u} defines the SLE associated to this $f$. This introduces an ambiguity in the procedure, which has been explored within the LQC approach in \cite{SLEs_ours}. In that work, only natural choices for $f$ within this framework were considered, whose support thus included the bounce of LQC. In this current investigation, we will consider test functions that exclude it.

\section{Effect of the bounce in SLE{\lowercase{s}}}\label{sec:SLEs no bounce}

In \cite{SLEs_ours} we have shown that there are two families of test functions that can be seen as natural choices for the smearing function within LQC, and that provide SLEs that are very insensitive to their particular form. Namely, we have found that for a test function supported around the bounce of LQC the resulting SLE does not qualitatively depend on its shape or support, as long as it is wide enough. In the case of a test function supported on the expanding branch only, from the bounce onward, in \cite{SLEs_ours} only the case of a steep (but smooth) step function was investigated, in order to fully retain the contributions coming from the bounce. In this case, the SLE remains insensitive to the size of the support as long as it is wide enough. 
The resulting power spectrum inherits this independence on the choice of test function, and coincidentally shows good agreement with the one of a second order adiabatic vacuum state.

In this section we explore the consequences of not including the bounce in the support of the test function. This way we will be able to study also the effect of the shape of the test function when supported only on the expanding branch away from the high-curvature regime. This will allow us to provide a comparison with an analogous classical scenario of an FLRW model with a period of kinetic dominance prior to inflation.

\begin{figure}[h!]
\begin{center}
\includegraphics[width=.6\textwidth]{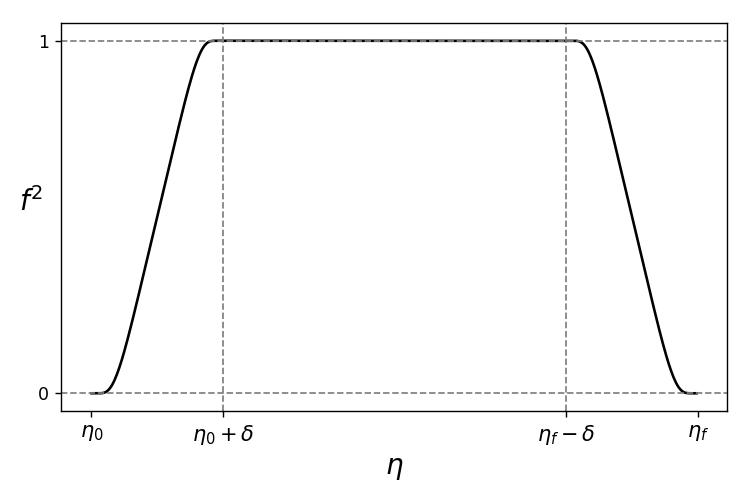}
\end{center}
\caption{The smooth step function defined in \eqref{eq:f2step}, represented in terms of its parameters: initial and final points, $\eta_i$ and $\eta_f$ respectively, and $\delta$, which controls the ramping up.}
\label{fig:f2}
\end{figure}

Let us start by considering the smooth step function $f^2$ plotted in Figure \ref{fig:f2}, supported in the interval $\eta \in \left[\eta_0,\eta_f\right]$, as defined in \cite{SLEs_ours}:
\begin{equation}\label{eq:f2step}
    f^2(\eta) = 
\begin{cases}
	S\left(\frac{\eta-\eta_0}{\delta}\pi\right) &\eta_0\leq \eta < \eta_0+\delta,\\
	1 & \eta_0+\delta\leq \eta \leq \eta_f-\delta,\\
	S\left(\frac{\eta_f-\eta}{\delta}\pi\right) &\eta_f-\delta < \eta \leq \eta_f,
\end{cases}
\end{equation}
where $\delta$ determines the ramping up, with a smaller $\delta$ resulting in a steeper step, and $S$ is the auxiliary function:
\begin{equation}
    S(x) = \frac{1-\tanh\left[\cot(x)\right]}{2}.
\end{equation}

\begin{figure}[h!]
\includegraphics[width=\textwidth]{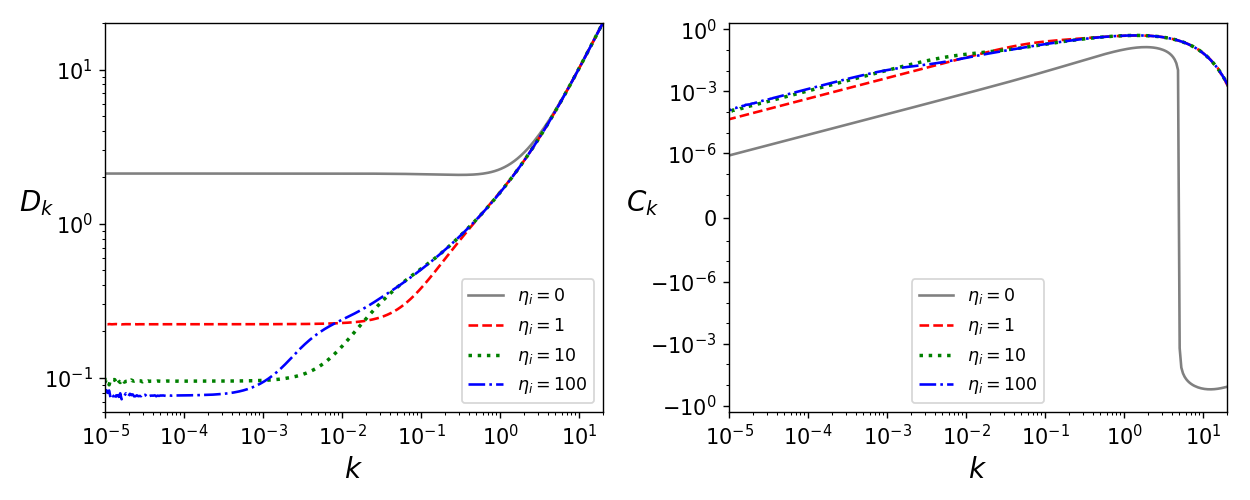}
\caption{Initial conditions in terms of $D_k$ and $C_k$, as constructed in \eqref{eq:CkDk}, for scalar modes at the bounce corresponding to the SLEs obtained with the window function \eqref{eq:f2step} covering the the expanding branch until the onset of inflation with starting points: $\eta_0=0$ (solid gray line), $\eta_0=1$ (dashed red line), $\eta_0=10$ (dotted green line) and $\eta_0=100$ (dotted-dashed blue line). The scale of $k$ is in Planck units. All computations were performed for a quadratic potential $V(\phi) = m^2\phi^2/2$, with $m=1.2\times 10^{-6}$ and with the value of the scalar field at the bounce fixed to $\phi_B = 1.225$ (toy value). For tensor modes, the resulting SLE at the bounce shows no significant qualitative differences.}
\label{fig:CD}
\end{figure}

Figure \ref{fig:CD} shows the initial conditions, parametrized through \eqref{eq:CkDk}, corresponding to the SLE obtained for scalar perturbations when considering the test function \eqref{eq:f2step}, with $\eta_0=0,1,10$ and $100$ Planck seconds after the bounce, with $\eta_f$ fixed at the onset of inflation, and for a sharp step of $\delta \sim 0.06$. The case of $\eta_0=0$ corresponds to the one analysed in \cite{SLEs_ours}. The effect of excluding the bounce is immediately noticed as soon as the support of the test function is moved one Planck second into the expanding branch. If we push the initial time further into the future, the change is gradually decreased, and for $\eta_0 = 100$ we see some convergence. The corresponding figure for tensor modes is omitted since the initial conditions are essentially the same, as discussed in \cite{SLEs_ours}. Within this family of test functions that exclude the bounce, we have also investigated the consequences of changing their shape. In all these cases, we find that, as the starting point moves further away from the bounce, the SLE becomes more insensitive to the shape of the test function. For this reason,  below, we will focus our comments on the four step functions defined above, as they already show the different qualitative behaviors one may obtain from different test functions in this scenario.

\begin{figure}[h!]
\includegraphics[width=\textwidth]{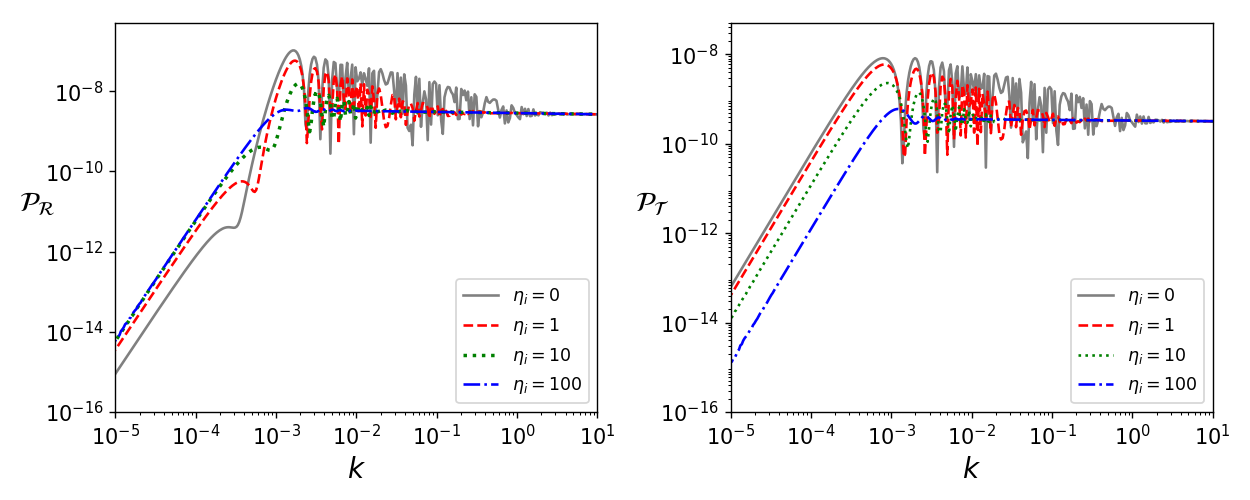}
\caption{Power spectra of the comoving curvature perturbation $\mathcal{P}_{\mathcal{R}}$ and tensor perturbation  $\mathcal{P}_{\mathcal{T}}$ corresponding to the SLEs obtained with the window function \eqref{eq:f2step} covering the the expanding branch until the onset of inflation with starting points: $\eta_0=0$ (solid gray line), $\eta_0=1$ (dashed red line), $\eta_0=10$ (dotted green line) and $\eta_0=100$ (dotted-dashed blue line). The scale of $k$ is in Planck units. All computations were performed for a quadratic potential $V(\phi) = m^2\phi^2/2$, with $m=1.2\times 10^{-6}$ and with the value of the scalar field at the bounce fixed to $\phi_B = 1.225$ (toy value).}
\label{fig:PS}
\end{figure}

Figure \ref{fig:PS} shows the corresponding primordial power spectra for scalar and tensor perturbations, computed through \eqref{eq:PS_RT}. Here, the effect of removing the bounce is evident. As the support of the test function is pushed further away from the high curvature regime, the oscillations in the power spectra are gradually dampened. 

\begin{figure}[h!]
\includegraphics[width=\textwidth]{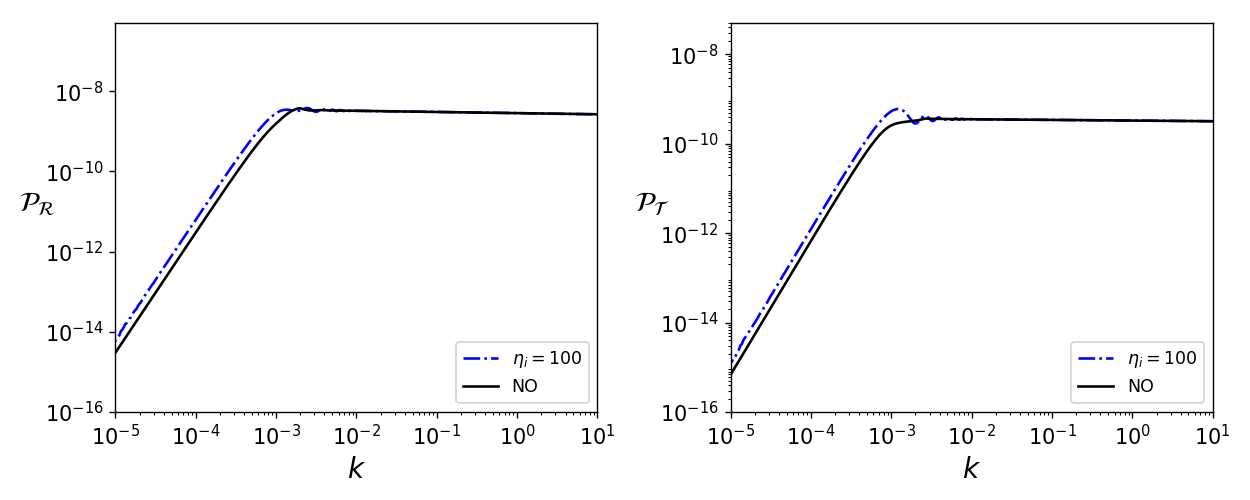}
\caption{Comparison between the power spectra of the comoving curvature perturbation $\mathcal{P}_{\mathcal{R}}$ and tensor perturbation  $\mathcal{P}_{\mathcal{T}}$ corresponding to the SLE obtained with $\eta_0=100$ (dotted-dashed blue line) and to the NO vacuum (solid black line). The scale of $k$ is in Planck units. All computations were performed for a quadratic potential $V(\phi) = m^2\phi^2/2$, with $m=1.2\times 10^{-6}$ and with the value of the scalar field at the bounce fixed to $\phi_B = 1.225$ (toy value).}
\label{fig:PSNO100}
\end{figure}

It is interesting to note that, in fact,  as Figure \ref{fig:PSNO100} shows, the power spectra are pushed towards those obtained from the non-oscillatory (NO) vacuum state defined in \cite{nonosc}, which is constructed by minimizing the oscillations in time of the power spectrum of perturbations for the whole expanding branch, including the bounce. We further note that the case where the support of the test function starts at $\eta_0=100$ will essentially correspond to that obtained by using the SLE as the vacuum state of primordial perturbations in a classical FLRW model with a period of kinetic dominance prior to inflation. However, for smaller $\eta_0$, SLEs show oscillations in $k$ at and below scales comparable to those of the curvature at that initial time. Then we  can conclude that the oscillations that appear in the power spectra when including the high curvature region (for instance the bounce of LQC) in the support of the test function open an interesting observational window. 

For completion, we added an appendix where we apply the SLE and NO vacuum prescriptions in a classical universe dominated by the kinetic energy of the scalar field. We discuss the situations in which they agree with the natural choice for vacuum state considered in \cite{Contaldi:2003zv}.

\section{Comparison between SLE and NO vacuum}\label{sec:SLEvsNO}

One remarkable property of the SLEs that is usually not explicitly proven for other vacua proposals is that they are of Hadamard type \cite{Olbermann2007,Niedermaier2020}. This guarantees that computations such as that of the expectation value of the renormalized stress-energy tensor will be well defined. On the other hand, the NO vacuum has only been proven to behave in the ultraviolet (UV) asymptotic regime as a high order adiabatic state, at least of fourth order \cite{nonosc_BeaDaniGuillermo}. Indeed, considering two adiabatic states of orders $n$ and $m$, one can compute the $\beta$ coefficient of the Bogoliubov transformation between the two:
\begin{equation}
\beta = i\left[ u^{n}_k \left(u^{ m}_k\right)^{\prime}-u^{ m}_k \left(u^{ n}_k\right)^{\prime}\right],
\end{equation}
and find that in the UV $|\beta|$ decays with $k^{-l-2}$, where $l = {\rm{min}}(n,m)$. In the case of the comparison of the NO vacuum with an nth-order adiabatic one, it was found that $|\beta|\sim k^{-2-n}$ at least up to $n=4$, which shows that the NO vacuum is the highest order one of the two. As a Hadamard type vacuum is an infinite order adiabatic state, this is an indication that the NO vacuum might be as well, though a stronger proof would be desirable. In this section, we will provide one, through a comparison with the (Hadamard-like) SLEs.

To simplify the comparison, let us write the UV expansions of both the SLE and the NO state as
\begin{equation}\label{eq:uF}
    u_k(\eta) \sim \frac{1}{\sqrt{2 F_k(\eta)}} e^{-i \int d\eta F_k(\eta)},
\end{equation}
where $\sim$ means the behavior in the large $k$ regime. The NO vacuum state has recently been analysed analytically in \cite{menava}. In particular, that work has found that the state admits the UV asymptotic expansion \eqref{eq:uF} with:
\begin{equation}\label{eq:F_NO}
    F_k^{NO}(\eta) = - {\rm Im}(h_k(\eta)),
\end{equation}
where
\begin{equation}
    kh_k^{-1}\sim i\left[1-\frac{1}{2k^2}\sum_{n=0}^{\infty}\left(\frac{-i}{2k}\right)^{n}\gamma_n \right],
\end{equation}
and the $\gamma_n$ coefficients are given by the iterative relation
\begin{equation}\label{eq:gamman}
\gamma_{n+1}=-\gamma_{n}^{\prime}+4s(\eta)\left[\gamma_{n-1}+\sum_{m=0}^{n-3}\gamma_m \gamma_{n-(m+3)}\right]-\sum_{m=0}^{n-1}\gamma_m \gamma_{n-(m+1)},
\end{equation}
with $\gamma_{0}=s(\eta)$ and $\gamma_{-n}=0$ for all $n>0$. With this expansion, we are able to compute the NO state up to any order in $1/k$ easily. Actually, one can check by direct inspection that 
\begin{equation}\label{eq:uFNO}
    F_k^{NO}(\eta) \sim k \left\{ 1 + \sum_{n\geq 1} \frac{(-1)^n}{k^{2n}} G_n(\eta)\right\}^{-1},
\end{equation}
where the $G_n$ are determined recursively by
\begin{equation}\label{eq:Gn}
    G_n(\eta) = \!\!\sum_{m,l\geq 0, m+l =n-1}\! \left\{\frac{1}{4} G_m G_l^{\prime\prime} -\frac{1}{8}G_m^{\prime} G_l^{\prime} +\frac{1}{2}s(\eta)G_m G_l\right\}-\frac{1}{2}\sum_{m,l\geq 1, m+l=n} G_m G_l,
\end{equation}
with $G_0=1$. Remarkably, in \cite{Niedermaier2020}, the SLEs are found to have the same asymptotic expansion \eqref{eq:uFNO}, regardless of the choice of the test function.

Therefore, the $\beta$ coefficients of the Bogoliubov transformation between the SLE and NO vacuum are identically zero in the UV. Thus, we conclude that the NO vacuum is of Hadamard type since it displays exactly the same short-distance structure as the SLEs. 

\section{Conclusion and discussion}\label{sec:conc}

SLEs have recently been proposed as a suitable choice for the vacuum state of perturbations in LQC \cite{SLEs_ours}, where the dependence of the state on the test function was explored. For that investigation, only test functions that included the bounce of LQC were analysed, as they are natural choices within this framework. In this work, we investigate the effect of pushing the support of the test function away from the bounce and indeed from the high curvature regime. In addition to offering a more complete analysis of SLEs within LQC, this allows us to disentangle the effects coming from quantum corrections to the dynamics, which are important in the high curvature regime, from those that arise from having only a period of kinetic dominance prior to inflation, which can be found also in classical inflationary models, and is not a direct consequence of the quantum nature of geometry.

We have found that whether the support of the test function includes the high curvature regime or not has a greater influence on the resulting SLE than any other parameter of the test function that has been studied previously. Indeed, in \cite{SLEs_ours}, we had already shown that as long as the support of the test function includes the high curvature regime and it is wide enough, the SLE is very insensitive to its shape and support. In this work, we have shown that as soon as the test function is pushed away from the bounce, the SLE suffers a big shift, which then converges as the test function is pushed further away from the high curvature regime. We have also found that in this case, when convergence with respect to the support is reached, the SLE is again insensitive to the shape of the test function. Furthermore, through the computation of the power spectra of perturbations at the end of inflation, we see that as the test function is shifted away from the high curvature region the oscillations found for lower modes (scales comparable to those of the curvature at initial time) are gradually dampened, and the spectra are pushed to those of the NO vacuum state introduced in \cite{nonosc}. Then it is safe to conclude that these oscillations are in fact a consequence of the corrections coming from LQC, which opens an interesting observational window into signatures from LQC in observations of the CMB.  For instance, the enhancement of power at super-Hubble scales in the primordial power spectrum of scalar perturbations has been proposed, together with large scale non-Gaussianities, as a mechanism to explain several anomalies in the CMB \cite{Agullo:2020,Agullo:2020b}. From this perspective, the power spectrum provided by SLEs prescription when including the bounce is physically relevant. On the other hand, the NO-like power spectra show a lack of power at large scales of primordial origin that can, on the one hand, alleviate some tensions in the CMB \cite{Ashtekar:2020,Ashtekar:2021}, and on the other hand, ease the trans-Planckian issues on these scenarios \cite{Brandenberger:2012,Ashtekar:2016}. However, a detailed analysis of all this requires a rigorous investigation that we leave for future work.

Finally, the fact that SLEs are proven to be Hadamard is a great advantage that most proposals don't enjoy. Typically, this property is difficult to prove explicitly. One strategy, that may be enough for practical purposes, is to compare a state with an adiabatic one of increasing (finite) order, and show, through the $\beta$ coefficients of the Bogoliubov transformation between the two states, that the state in question is always of higher order than the adiabatic one considered. This shows that it is at least a very high order adiabatic state, and since  a Hadamard state is an adiabatic state of infinite order, then most likely so is the proposed state. However, we now have a  family of states, namely SLEs, that are explicitly Hadamard. Therefore, the $\beta$ coefficients of the transformation between any Hadamard state and any SLE should decrease faster than any power of the wave number. We have applied this reasoning to the NO vacuum state, that had previously been shown to be at least of fourth order \cite{nonosc_BeaDaniGuillermo}.  We find that, in the ultraviolet limit of large wave numbers, the asymptotic expansion that the NO vacuum satisfies (found in \cite{menava}) agrees exactly with that of the SLE \cite{Niedermaier2020} (no matter the test function chosen to define it). As a consequence, the $\beta$ coefficients of the transformation between the two will be identically zero in the ultraviolet. In other words, the NO vacuum has the same short-distance structure than the SLEs, which proves that the NO vacuum is of Hadamard type as well.

\acknowledgements

The authors would like to thank Beatriz  Elizaga  Navascu\'es and Guillermo Mena Marug\'an for useful discussions. This work is supported by the Spanish Government through the projects FIS2017-86497-C2-2-P and PID2019-105943GB-I00 (with FEDER contribution). R.~N. acknowledges financial support from Funda\c{c}\~{a}o para a Ci\^{e}ncia e a Tecnologia (FCT) through the research grant SFRH/BD/143525/2019. J. O. acknowledges the Operative Program FEDER2014-2020 and the Consejer\'ia de Econom\'ia y Conocimiento de la Junta de Andaluc\'ia.

\appendix

\section*{NO and SLEs vacuum in kinetically dominated universes}\label{app}

 In General Relativity, an early universe dominated by the kinetic energy of the inflaton field is well  described by a spacetime coupled to a perfect fluid with constant equation of state $P=\omega \rho$ with $\omega=1$, namely, stiff matter. The scale factor of the spacetime behaves as $a(\eta)=\sqrt{\eta}$ in conformal time $\eta$. The equation of motion of scalar and tensor perturbations is also well described by Eq. \eqref{eq:eom_uk}, with
\begin{equation}
    s(\eta)=\frac{1}{4\eta^2}.  
\end{equation}
Any complex solution can be written in terms of well-known Hankel functions $W_k(\eta)=\sqrt{\frac{\pi \eta}{4}}\bar H^{(1)}_0(k\eta)$ and its complex conjugate. 

Actually, that is the natural initial state that one adopts in a kinetically dominated classical universe \cite{Contaldi:2003zv}. Let us show that $|W_k(\eta)|^2$ has an asymptotic expansion for large $k$ of the form of the SLEs, namely, it is of the form of Eq. \eqref{eq:uFNO} --the phase of $W_k(\eta)$ is determined by $|W_k(\eta)|$ via Eq. \eqref{eq:uF} which is itself a consequence of the normalization of this basis of complex solutions. Using the asymptotic expansion of Hankel functions
\begin{equation}
H_{\nu}^{(1)}(z) \sim\left(\frac{2}{\pi z}\right)^{1 / 2} e^{i\left(z-\frac{1}{2} \nu \pi-\frac{1}{4} \pi\right)} \sum_{m=0}^{\infty}i^{m} \frac{a_{m}(\nu)}{z^{m}},
\end{equation}
with 
\begin{equation}
    a_{m}(\nu)=\frac{\left(4 \nu^{2}-1^{2}\right)\left(4 \nu^{2}-3^{2}\right) \cdots\left(4 \nu^{2}-(2 m-1)^{2}\right)}{m ! 8^{m}},
\end{equation}
one obtains
\begin{equation}
[F^{W_k}(\eta)]^{-1}=|W_{k}(\eta)|^2\sim\frac{1}{2k} \left(\sum_{m=0}^{\infty}i^{m} \frac{a_{m}(0)}{(k\eta)^{m}}\right)\left(\sum_{n=0}^{\infty} (-i)^{n} \frac{a_{n}(0)}{(k\eta)^{n}}\right),
\end{equation}
or more explicitly,
\begin{equation}
[F^{W_k}(\eta)]^{-1}\sim\frac{1}{2k} \left(1-\frac{1}{8 {k}^{2} {\eta}^{2}}+\frac{27}{128 {k}^{4} {\eta}^{4}}-\frac{1125}{1024 {k}^{6} {\eta}^{6}}+\cdots\right).
\end{equation}
On the other hand, using the recursion relation in Eq. \eqref{eq:Gn}, one obtains,
\begin{equation}
    G_1(\eta)=\frac{1}{8\eta^2}, \quad G_2(\eta)=\frac{27}{128\eta^4},\quad G_3(\eta)=\frac{1125}{1024\eta^6},\quad \cdots
\end{equation}

The asymptotic behavior at large $k$'s exactly agrees with the one of SLEs. Hence, we can claim that the state $W_k(\eta)$ is of Hadamard type. 

Now, let us see how this state is related to the SLE and the NO vacuum.  We start with the SLE prescription. Let us consider an arbitrary state $\omega$, whose modes will be denoted by $T^\omega_k$. This state will be related to the fiducial state defined by $W_k$ via a Bogoliubov transformation
 \begin{align}
 T^\omega_k=\alpha(k)W_k+\beta(k)\bar{W}_k.
 \end{align}
 Then, the smeared energy density of the mode $T^\omega_k$ is given by
\begin{align}
    E(T^{\omega}_k)=E(W_k)(1+2\beta^2(k)) +2\beta(k)\sqrt{1+\beta^2(k)}\mathcal{R}[e^{i\arg[\alpha(k)]} D(W_k)],
\end{align}
with
\begin{equation}
    E(W_k) = \frac{1}{2} \int_{\eta_0}^{\eta_f} d\eta\,f^2(\eta) \frac{k^2}{\eta} \left[\left\lvert H^{(1)}_1(k\eta)\right\rvert^2 + \left\lvert H^{(1)}_0(k\eta)\right\rvert^2\right],  
\end{equation}
and
\begin{equation}
   D(W_k) = \frac{1}{2} \int_{\eta_0}^{\eta_f} d\eta\,f^2(\eta) \frac{k^2}{\eta} \left[\left( H^{(1)}_1(k\eta) \right)^2 +\left( H^{(1)}_0(k\eta) \right)^2\right],
\end{equation} 
 Here we are calling $[\eta_0,\eta_f]$ the support of $f$, and we have used the property $\frac{d}{dx} H^{(1)}_0(x)=-H^{(1)}_1(x)$.
We obtain the  smallest value for $E(T^{\omega}_k)$ by choosing $\arg[\alpha(k)]$ such that $e^{i\arg[\alpha(k)]} D(W_k)=-|D(W_k)|$ and $\beta(k)\geq0$. If we define $\delta(k)=\beta(k)/\sqrt{1+\beta(k)^2}\in[0,1)$, then,  no matter the modes $T^\omega_k$ considered, we can write
\begin{align}
   E(T^{\omega}_k)\geq\mathcal{R}_k(\eta_f,\eta_0)E(W_k),
\end{align}
where
\begin{align}
   \mathcal{R}_k(\eta_f,\eta_0)=\left\{1+\frac{2\delta(k)}{1-\delta(k)^2}\left[\delta(k) -\frac{|D(W_k)|}{E(W_k)}\right]\right\}.
\end{align}
We have found that for test functions well approximated by a (smooth) step function with support in $[\eta_0,\eta_f]$ and in the limit $\eta_f\to\infty$, the ratio $|D(W_k)|/E(W_k)$ behaves as
\begin{equation}
\frac{|D(W_k)|}{E(W_k)} \simeq\frac{1}{2\pi(k\eta_0)^2}.
\end{equation}
Therefore,  in the limit $(k\eta_0)\gg 1$ we obtain $\mathcal{R}_k(\eta_f,\eta_0)\geq1$, and then we always have $E(T^{\omega}_k)\geq E(W_k)$. We thus conclude that the state defined by the modes $W_k$ is a SLE provided that $(k\eta_0)\gg 1$.

On the other hand, for the NO vacuum, if we follow the prescription in \cite{nonosc}, it can be implemented as the solution $v_{k}^{(NO)}(\eta)$ that minimizes the integral
\begin{equation}
    I(v_k)=\int_{\eta_0}^{\eta_f}d\eta \left|\partial_\eta|v_{k}(\eta)|^2\right|.
\end{equation}
Namely, $I(v_k)\geq I(v_k^{(NO)})$. If we write 
\begin{equation}
    v_k^{(NO)}(\eta)=\underline{\alpha}(k) W_k(\eta)+\underline{\beta}(k) \bar W_k(\eta),
\end{equation}
we have that
\begin{equation}
    |v_k^{(NO)}(\eta)|^2= J_k(\eta)|W_k(\eta)|^2,
\end{equation}
with
\begin{equation}
    J_k(\eta)=1+2|\underline{\beta}(k)|^2+2|\underline{\beta}(k)|\sqrt{1+|\underline{\beta}(k)|^2}\cos\left\{\arg[\beta(k)]-\arg[W_k(\eta)]\right\}.
\end{equation}
One can see that $J_k(\eta)\geq 0$. Now, the integral above can be written as
 \begin{equation}\label{eq:no-int}
    I(v_k^{(NO)})=\int_{\eta_0}^{\eta_f}d\eta \left|J_k(\eta)\partial_\eta|W_{k}(\eta)|^2+|W_{k}(\eta)|^2\partial_\eta J_k(\eta)\right|.
\end{equation}
In the cases in which $\partial_\eta|W_{k}(\eta)|^2=0$ one can easily see that the minimum of this integral is reached iff $\underline\beta(k)=0$. However, as in the SLEs, by looking at the asymptotic expansion of the solutions $W_k(\eta)$ in terms of the Hankel functions for large argument, namely, for $k\eta\gg 1$ one can see that
\begin{equation}
    |W_k(\eta)|^2=\frac{1}{2k}\left(1+{\mathcal O}\left[(k\eta)^{-2}\right]\right),\quad \arg[W_k(\eta)]=-k\eta\left(1+{\mathcal O}\left[(k\eta)^{-2}\right]\right)
\end{equation}
Therefore, $\partial_\eta|W_{k}(\eta)|^2$ decreases sufficiently fast for  $k\eta\gg1$. Besides, one can see that the leading and subleading contributions in the integrand in Eq. \eqref{eq:no-int} in the limit $k\eta\gg1$ is
\begin{equation}
    \left||\underline{\beta}(k)|\sqrt{1+|\underline{\beta}(k)|^2}\sin(\arg(\beta)+k\eta)+{\mathcal O}(k\eta)^{-3}\right|. 
\end{equation}
Therefore, in the limit $\eta_f\to\infty$, the integral $I(v_k^{(NO)})$ will diverge unless $\beta(k)=0$ for all $k$'s. 

In summary, both the SLE and NO prescriptions agree in the regimes discussed in this Appendix. Concretely, in the situations in which the smearing is carried over sufficiently large values of conformal times.

\bibliography{SLEvsNObib}

\end{document}